\documentclass[runningheads,openany]{svmult}
\usepackage{cite}
\usepackage[T1]{fontenc}
\usepackage{palatino}
\usepackage{euler}
\usepackage{graphicx}  
\usepackage{physprbb}
\usepackage{bg2002}
\usepackage{epsfig}

\renewcommand{\deg}{^{\circ}}

\newcommand{\pip}[1]{\left( #1 \right)} 

\newcommand{\eqnb}{\begin{equation}}
\newcommand{\eqne}{\end{equation}}

\renewcommand{\question}[1]{}
\newcommand{\NSt}{{\mbox{\scriptsize\it NS}}}
\newcommand{\St}{{\mbox{\scriptsize\it S}}}

\begin{document}
\setcounter{page}{27}


\let\I\i
\def\i{\mathrm{i}}
\def\e{\mathrm{e}}
\def\d{\mathrm{d}}
\def\half{{\textstyle{1\over2}}}
\def\thalf{{\textstyle{3\over2}}}
\def\h{{\scriptscriptstyle{1\over2}}}
\def\th{{\scriptscriptstyle{3\over2}}}
\def\fh{{\scriptscriptstyle{5\over2}}}
\def\vec#1{\mbox{\boldmath$#1$}}
\def\svec#1{\mbox{{\scriptsize \boldmath$#1$}}}
\def\oN{\overline{N}}
\def\ttimes{{\scriptstyle \times}}
\def\bm#1{{\pmb{\mbox{${#1}$}}}}

\def\CG#1#2#3#4#5#6{C^{#5#6}_{#1#2#3#4}}
\def\threej#1#2#3#4#5#6{\left(\begin{array}{ccc}
    #1&#2&#3\\#4&#5&#6\end{array}\right)}

\title*{%
Phenomenological Schwinger-Dyson approach to \\
$\eta$-$\eta'$ mass matrix
\thanks{Talk delivered by
D. Klabu\v{c}ar.
}}

\author{%
D. Kekez$^{a}$ and
D. Klabu\v{c}ar$^{b}$
}\institute{%
$^a${Rudjer Bo\v{s}kovi\'{c} Institute, P.O.B. 180, 10002 Zagreb, Croatia}\\
$^b${Physics Department, Faculty of Science, Zagreb University}
}

\authorrunning{D. Kekez and D. Klabu\v{c}ar}
\titlerunning{Phenomenological Schwinger-Dyson approach to $\eta$-$\eta'$ mass matrix}


\maketitle


\begin{abstract}
It is reviewed pedagogically how a very successful description of the 
$\eta$--$\eta^\prime$ mass matrix can be achieved in the 
consistently coupled Schwinger-Dyson and Bethe-Salpeter 
approach in spite of the limitations of the ladder approximation.
This description is in agreement with both phenomenology 
and lattice results.

\end{abstract}

\noindent
The Schwinger-Dyson and Bethe-Salpeter (SD-BS) approach is the 
bound-state approach that is chirally well-behaved. (For example, see 
Ref. \cite{Roberts:2000hi,Alkofer:2000wg,Roberts:2000aa} 
for recent reviews, and references therein 
for various phenomenological and other applications of SD-BS approach.)
Therefore, among the bound-state approaches it is probably the most 
suitable one to treat the light pseudoscalar mesons.
One solves the Schwinger-Dyson (SD) equation for dressed propagators
of the light $u, d$ and $s$ quarks using a phenomenologically
successful interaction. These light-quark propagators are then 
employed in consistent solving of Bethe-Salpeter (BS) equations for
various quark-antiquark ($q\bar q$) relativistic bound states. 
Namely, in both SD and BS equations we employ the same, ladder
approximation, and the same interaction due to an effective, 
{\it dressed} gluon exchange. In the chiral limit (and {\it close} 
to it), light pseudoscalar ($P$) meson $q\bar q$ bound states 
($P=\pi^{0,\pm}, K^{0,\pm}, \eta$) 
then {\bf simultaneously} manifest themselves also as 
({\it quasi-})Goldstone bosons of dynamically broken chiral symmetry.
This resolves the dichotomy ``$q\bar q$ bound state {\it vs.}
Goldstone boson", enabling one to work with the mesons as 
explicit $q\bar q$ bound states (for example, in Refs.
\cite{Kekez:1996az,Klabucar:1997zi,Kekez:1998xr,Kekez:1998rw,Kekez:2000aw,Klabucar:2000me,Kekez:2001ph,Bistrovic:1999dy}) 
while reproducing (even analytically, in the chiral limit) the famous 
results of the axial anomaly for the light pseudoscalar mesons, namely 
the amplitudes for $P\rightarrow\gamma\gamma$ and 
$\gamma^\star \rightarrow P^0 P^+ P^-$ \cite{Alkofer:1995jx}.
This is unique among the bound state approaches -- for example, see
Refs. \cite{Roberts:2000hi,Kekez:1998xr,Alkofer:1995jx} 
and references therein. Nevertheless, one keeps the advantage 
of bound state approaches that from the $q\bar q$ substructure
one can calculate many important quantities (such as the pion decay 
constant $f_\pi$) which are just parameters in most of other chiral 
approaches to the light-quark sector. The description 
\cite{Klabucar:1997zi,Kekez:2000aw,Klabucar:2000me,Kekez:2001ph}
of $\eta$--$\eta'$ complex is especially noteworthy, as it is 
very successful in spite of the limitations of the SD-BS approach
in the ladder approximation.

For the description of $\eta$ and $\eta'$, the crucial issues are 
the meson mixing and construction of physical meson states. They are 
formulated in Refs. \cite{Klabucar:1997zi,Kekez:2000aw,Klabucar:2000me} 
for the SD-BS approach, where solving of appropriate BS equations 
yields the eigenvalues of the squared masses, 
$M_{u\bar{u}}^2,M_{d\bar{d}}^2,M_{s\bar{s}}^2$ and $M_{u\bar{s}}^2$,
of the respective quark-antiquark bound states $|u\bar{u}\rangle , 
|d\bar{d}\rangle , |s\bar{s}\rangle$ and $|u\bar{s}\rangle$.
The last one is simply the kaon, and $M_{u\bar{s}}$ is its mass $M_K$.
Nevertheless, the first three do not correspond to any physical 
pseudoscalar mesons.
Thus, $M_{u\bar{u}}^2,M_{d\bar{d}}^2,M_{s\bar{s}}^2$ do not automatically 
represent any physical masses, although the mass matrix 
(actually, to be precise, the non-anomalous part of the mass matrix) is
simply ${\hat M}^2_{NA} = \mbox{\rm diag} (M_{u\bar{u}}^2,M_{d\bar{d}}^2,M_{s\bar{s}}^2)$
in the basis $|q\bar{q}\rangle, (q=u,d,s)$. However, the flavor SU(3) quark 
model leads one to recouple these states into the familiar octet-singlet basis 
of the zero-charge subspace of the light unflavored pseudoscalar mesons:
\begin{eqnarray}
        |\pi^0\rangle
        &=&
        \frac{1}{\sqrt{2}} (|u\bar{u}\rangle - |d\bar{d}\rangle)~,
\label{pi0def}
        \\
        |\eta_8\rangle
        &=&
        \frac{1}{\sqrt{6}} (|u\bar{u}\rangle + |d\bar{d}\rangle
                                            -2 |s\bar{s}\rangle)~,
\label{eta8def}
        \\
        |\eta_0\rangle
        &=&
        \frac{1}{\sqrt{3}} (|u\bar{u}\rangle + |d\bar{d}\rangle
                                             + |s\bar{s}\rangle)~.
\label{eta0def}
        \end{eqnarray}
With $|u\bar{u}\rangle$ and $|d\bar{d}\rangle$ being practically
chiral states as opposed to a significantly heavier $|s\bar{s}\rangle$,
Eqs.~(\ref{pi0def}--\ref{eta0def}) do not define the octet and singlet states 
of the exact SU(3) flavor symmetry, but the {\it effective} octet and 
singlet states. However, as pointed out by Gilman and Kauffman \cite{Gilman:1987ax}  
(following Chanowitz, their Ref. [8]), in spite of this flavor symmetry breaking by the 
$s$ quark, these equations still implicitly assume nonet symmetry in
the sense that the same states $|q\bar{q}\rangle$ ($q=u,d,s$)
appear in both the octet member $\eta_8$ (\ref{eta8def}) and the singlet
$\eta_0$ (\ref{eta0def}). 
Nevertheless, in order to avoid the U$_A$(1) problem, this symmetry must 
ultimately be broken by gluon anomaly at least at the level of the 
masses of pseudoscalar mesons.

In the basis (\ref{pi0def}--\ref{eta0def}),
the non-anomalous part of the (squared-)mass 
matrix of $\pi^0$ and etas is
\begin{equation}
{\hat M}^2_{NA} =
\left( \begin{array}{ccl} M_{\pi}^2 & 0 & 0 \\
                          0 & M_{88}^2 & M_{80}^2\\
                          0 & M_{08}^2 & M_{00}^2
        \end{array} \right)~.
\label{M2NA}
\end{equation}
The $\eta_8$ ``mass" $M_{88}\equiv M_{\eta_8}$ can be related to the kaon
mass through the Gell-Mann--Okubo (GMO) relation, although the kaon does 
not appear in this scheme as it obviously cannot mix with $\pi^0$ and etas,
since it is strangely flavored.
Equation (\ref{M2NA}) shows that also the isospin $I=1$ state $\pi^0$ decouples 
from any mixing with the $I=0$ states $\eta_8$ and $\eta_0$, thanks to our 
working in the isospin limit throughout. Therefore, we are concerned only
with the diagonalization of the $2\times 2$ submatrix in the subspace of etas 
in order to find their physical masses and corresponding $q\bar q$ content.
In the isospin limit, obviously $M_{u\bar{u}} = M_{d\bar{d}}$, which we 
then can strictly identify with our model $\pi^0$ mass $M_{\pi}$. Since
in this model we can also calculate 
$M_{s\bar{s}}^2 = \langle s\bar{s} | {\hat M}^2_{NA} | s\bar{s} \rangle$,
this gives us our calculated entries in the mass matrix:
\begin{equation}
M_{88}^2 \equiv \langle \eta_8 | {\hat M}^2_{NA} |\eta_8 \rangle
\equiv M_{\eta_8}^2 = \frac{2}{3} (M_{s\bar{s}}^2 + \frac{1}{2}M_{\pi}^2) \, ,
\end{equation}
\begin{equation}
M_{80}^2 \equiv \langle \eta_8 | {\hat M}^2_{NA} |\eta_0 \rangle
= M_{08}^2 = \frac{\sqrt{2}}{3} ( M_{\pi}^2 - M_{s\bar{s}}^2 ) \, ,
\end{equation}
\begin{equation}
M_{00}^2 \equiv \langle \eta_0| {\hat M}^2_{NA} |\eta_0 \rangle
= \frac{2}{3} (\frac{1}{2}M_{s\bar{s}}^2 + M_{\pi}^2) \, .
\end{equation}
The last one, $M_{00}$, is the {\it non-anomalous part} of the 
$\eta_0$ ``mass" $M_{\eta_0}$. Namely, all the model masses 
$M_{q\bar q'}$ ($q, q' = u,d,s$) and corresponding $q\bar q'$ 
bound state amplitudes are 
obtained in the ladder approximation, and thus (irrespective
of what the concrete model could be) with an interaction kernel 
which cannot possibly capture the effects of gluon anomaly.
Fortunately, the large $N_c$ expansion indicates that the 
leading approximation in that expansion describes the bulk
of main features of QCD.
The gluon anomaly is suppressed as $1/N_c$ and is viewed as a 
perturbation in the large $N_c$ expansion. It is coupled {\it only} 
to the singlet combination $\eta_0$ (\ref{eta0def}); only the $\eta_0$ 
mass receives, from the gluon anomaly, a contribution which, unlike 
quasi-Goldstone masses $M_{q\bar q'}$'s comprising ${\hat M}^2_{NA}$, 
does {\it not} vanish in the chiral limit.
As discussed in detail in Sec. V of Ref. \cite{Klabucar:1997zi}, 
in the present bound-state context it is thus best to adopt the 
standard way (see, e.g., Refs. \cite{Miransky,Donoghue:dd})
to {\it parameterize} the anomaly effect. We thus break the $U_A(1)$
symmetry, and avoid the $U_A(1)$ problem, by shifting the $\eta_0$ 
(squared) mass by an amount denoted by $3\beta$ (in the notation of 
Refs. \cite{Kekez:2000aw,Klabucar:2000me}). The complete mass matrix 
is then ${\hat M}^2 = {\hat M}^2_{NA} +  {\hat M}^2_A$, where 
\begin{equation}
{\hat M}^2_A = 
\left( \begin{array}{ccl} 0 & \, 0 & \, 0 \\
                          0 & \, 0 & \, 0 \\
                          0 & \, 0 & 3\beta
        \end{array} \right)~,
\label{M2A}
\end{equation}
and where the value of the anomalous $\eta_0$ mass shift $3\beta$ 
is related to the topological susceptibility of the vacuum,
but in the present approach must be treated as a parameter to 
be determined outside of our bound-state model, i.e., fixed by
phenomenology or taken from the lattice calculations \cite{Alles:1996nm}.

We could now go straight to the nonstrange-strange (NS-S) basis, but 
before doing this, it may be instructive to rewrite for a moment the 
matrix (\ref{M2A})  in the flavor, $|q\bar q\rangle$ basis, where
\begin{equation}
{\hat M}^2_A = \beta
\left( \begin{array}{ccl} 1 & 1 & 1 \\
                          1 & 1 & 1 \\
                          1 & 1 & 1
        \end{array} \right)~,
\label{M2Aqq}
\end{equation}
since for some readers it may be the best place to introduce the effect of
flavor symmetry breaking. Namely, Eq. (\ref{M2Aqq}) tells us that due to the gluon
anomaly, there are transitions $|q\bar q\rangle \to |q' \bar q' \rangle $;
$q, q' = u, d, s$. However, the amplitudes for the transition from, and into,
light $u\bar u$ and $d\bar d$ pairs need not be the same as those for the
significantly more massive $s\bar s$.
The modification of the anomalous mass matrix (\ref{M2Aqq}) which allows
for possible effects of the breaking of the SU(3) flavor symmetry is then

\begin{equation}
{\hat M}^2_A = \beta
\left( \begin{array}{ccl} 1 & 1 & X \\
                          1 & 1 & X \\
                          X & X & X^2
        \end{array} \right)~.
\label{M2AqqX}
\end{equation}
There are arguments \cite{Kekez:2000aw,Klabucar:2000me}, supported by 
phenomenology, that the transition suppression is estimated well by
the nonstrange-to-strange ratio of respective constituent masses, 
$X\approx {\cal M}_u/{\cal M}_s$, or, as commented below, of 
respective decay constants, $X \approx f_\pi/f_{s\bar s}$.

After adding the anomalous contribution (\ref{M2A}) to Eq. (\ref{M2NA}),
pion still remains decoupled and we obviously still can 
restrict ourselves to $2\times 2$ submatrix in 
the subspace of etas.
However, when dealing with quark degrees of freedom when the symmetry 
between the nonstrange ({\it NS}) and strange ({\it S}) sectors is 
broken as described above, the most suitable basis for that subspace 
is the so-called {\it NS}-{\it S} basis:
        \begin{eqnarray}
        |\eta_\NSt\rangle
        &=&
        \frac{1}{\sqrt{2}} (|u\bar{u}\rangle + |d\bar{d}\rangle)
  = \frac{1}{\sqrt{3}} |\eta_8\rangle + \sqrt{\frac{2}{3}} |\eta_0\rangle~,
\label{etaNSdef}
        \\
        |\eta_\St\rangle
        &=&
            |s\bar{s}\rangle
  = - \sqrt{\frac{2}{3}} |\eta_8\rangle + \frac{1}{\sqrt{3}} |\eta_0\rangle~.
\label{etaSdef}
        \end{eqnarray}
The $\eta$--$\eta^\prime$ mass matrix in this basis is
\eqnb
{\hat M}^2 =
             \pip{
                \begin{array}{ll}
         M_{\eta_{NS}}^2    &  M_{\eta_{S}\eta_{NS}}^2 \\
            M_{\eta_{NS}\eta_{S}}^2 &    M_{\eta_{S}}^2
                \end{array}
        }
     =
       \pip{
                \begin{array}{ll}
      M_{u\bar{u}}^2 + 2 \beta  & \quad \sqrt{2} \beta X \\
        \,  \sqrt{2} \beta X    & M_{s\bar{s}}^2 + \beta X^2
                \end{array}
        }
\begin{array}{c} \vspace{-2mm} \longrightarrow \\ \phi \end{array}
        \pip{
                \begin{array}{ll}
                        M_\eta^2        & 0 \\
                        0               & M_{\eta'}^2
                \end{array}
        },
        \label{M2_NS-S}
\eqne
where the indicated diagonalization is given by
the {\it NS--S} mixing relations{\footnote{The effective-singlet-octet 
mixing angle $\theta$, defined by analogous relations where 
$ \eta_\NSt \to \eta_8, \eta_\St \to \eta_0, \phi \to \theta$, 
is related to the {\it NS--S} mixing angle $\phi$ as 
$\theta = \phi - \arctan \sqrt{2} =  \phi - 54.74\deg$.
The relation between our approach and the two-mixing-angle scheme 
is clarified in the Appendix of Ref. \cite{Kekez:2000aw}.}}
\begin{equation}
|\eta\rangle = \cos\phi |\eta_\NSt\rangle
             - \sin\phi |\eta_\St\rangle~,
\,\,\,\,\,\,\,
|\eta^\prime\rangle = \sin\phi |\eta_\NSt\rangle
             + \cos\phi |\eta_\St\rangle~,
\label{eqno3}
\end{equation}
rotating $\eta_\NSt,\eta_\St$ to the mass eigenstates $\eta, \eta'$.  
Now the {\it NS--S} mass matrix (\ref{M2_NS-S}) tells us that due to the gluon anomaly,
there are transitions $| \eta_\NSt \rangle \leftrightarrow | \eta_\St \rangle$. 
However, the amplitude for the transition from, and into, $\eta_\NSt$,
need not be the same as those for the more massive $\eta_\St$. 
The role of the flavor-symmetry-breaking factor $X$ is to allow for that 
possibility. As remarked little earlier, there are arguments 
\cite{Kekez:2000aw,Klabucar:2000me}, supported by 
phenomenology, that the transition suppression is estimated well by
the nonstrange-to-strange ratio of respective quark constituent masses, 
${\cal M}_u $ and ${\cal M}_s $. 
Due to the Goldberger-Treiman relation, this ratio is 
essentially equal \cite{Klabucar:1997zi,Kekez:2000aw,Klabucar:2000me} to
the ratio of $\eta_\NSt$ and $\eta_\St$ pseudoscalar decay constants   
$f_{\eta_\NSt}=f_\pi$ and $f_{\eta_\St}=f_{s\bar s}$ which are 
calculable in the SD-BS approach. Same as ${\cal M}_u$ and ${\cal M}_s$, 
they were found in our earlier papers, especially 
\cite{Klabucar:1997zi,Kekez:2000aw,Klabucar:2000me}.
In other words, we can estimate the flavor-symmetry-breaking suppression factor as 
$X\approx {\cal M}_u/{\cal M}_s$, or equivalently, as
$X \approx f_\pi/f_{s\bar s}$. 
Our model results ${\cal M}_u/{\cal M}_s = 0.622$
and $f_\pi/f_{s\bar s} = 0.689$
are in both cases reasonably close to $X_{exp} \approx 0.78$ extracted 
phenomenologically \cite{Kekez:2000aw,Klabucar:2000me} from the 
{\it empirical} mass matrix ${\hat m}^2_{exp}$ featuring experimental 
pion and kaon masses, or, after diagonalization, $\eta$ and $\eta'$ masses 
-- see Eq. (7) in Ref. \cite{Klabucar:2000me}. (In our model calculations
below, we use $X = f_\pi/f_{s\bar s} = 0.689$ to show the robustness of 
our approach to slight variations. 
Namely, our earlier works \cite{Kekez:2000aw,Klabucar:2000me,Klabucar:2001gr} 
mostly presented results based on slightly different values of $X$
obtained with the help of ratios of $\gamma\gamma$ amplitudes, which
is yet another, but again related way of obtaining $X$.)

In our present notation, capital $M_a$'s denote the calculated, model
pseudoscalar masses, whereas lowercase $m_a$'s denote the corresponding 
empirical masses.
The empirical mass matrix ${\hat m}^2_{exp}$ can be obtained from the
calculated, model one in Eq. (\ref{M2_NS-S}) by {\it i)} obvious 
substitutions $M_{u\bar u} \equiv M_\pi \rightarrow m_\pi$ and 
$M_{s\bar s} \rightarrow m_{s\bar s}$, and {\it ii)} by noting that 
$m_{s\bar s}$, the ``empirical" mass of the unphysical $s\bar s$ 
pseudoscalar bound state, is given in terms of masses of physical 
particles as $m_{s\bar s}^2 = 2 m_K^2 - m_\pi^2$ due to the 
Gell-Mann-Oakes-Renner (GMOR) relation. Since $M_{u\bar u}$, obtained 
by solving the BS equation, is identical to our model pion mass $M_\pi$, 
it was fitted to the empirical pion mass $m_\pi$, {\it e.g.}, in
Ref. \cite{Kekez:2000aw}. 
Similarly, 
$M_{u\bar s} \equiv M_K$ is fitted to the empirical kaon mass $m_K$. 
Therefore we also have  $M_{s\bar s}^2 \approx 2 m_K^2 - m_\pi^2$, since   
our model has good chiral behavior and also satisfies the GMOR relation, 
thanks to which we have
$M_{s\bar s}^2 = 2 M_K^2 - M_\pi^2$ in a very good approximation.
We thus see that in our model mass matrix, the parts stemming
from its {\it non-anomalous} part ${\hat M}^2_{NA}$ (\ref{M2NA}) 
are already close to the corresponding parts in ${\hat m}^2_{exp}$.
We can thus expect a good overall description of the masses in
$\eta$ and $\eta'$ complex. We now proceed to verify this expectation.

The {\it anomalous} entry $\beta$ is fixed phenomenologically 
to be $\beta_{exp} \approx 0.28$ GeV$^2$, 
along with $X_{exp} \approx 0.78$,
by requiring that trace and determinant of ${\hat m}^2_{exp}$ 
have their experimental values. But, this can be done also with
our calculated, model mass matrix ${\hat M}^2$. Requiring 
that the empirical value of the trace $m_\eta^2 + m_{\eta'}^2 
\approx 1.22$ GeV$^2$ be fixed by Eq. (\ref{M2_NS-S}), yields
\begin{equation}
\beta = \frac{1}{2+X^2} [(m_\eta^2 + m_{\eta'}^2)_{exp} -
(M_{u\bar u}^2 + M_{s\bar s}^2) ]
\end{equation}
where $X=f_\pi/f_{s\bar s} = 0.689$ and $M_{u\bar u}=0.1373$ GeV and
$M_{s\bar s} = 0.7007$ GeV are now our model-calculated \cite{Kekez:2000aw} 
quantities, giving us $\beta = 0.286$ GeV$^2$. 
Since $(M_{u\bar u}^2 + M_{s\bar s}^2) = 2m_K^2$ holds to a very good
approximation, our approach satisfies well the first equality (from the
matrix trace) in
\begin{equation}
2 \beta + \beta X^2 = m_\eta^2 + m_{\eta'}^2 - 2 m_K^2 =
\frac{2 N_f}{f_\pi^2} \, \chi \, ,
\label{WittenVenez}
\end{equation}
where the second equality is the Witten-Veneziano (WV) 
formula \cite{WV}, with $\chi$ being the topological susceptibility 
of the pure Yang-Mills gauge theory.  Our model values of 
$X$ and $\beta$ ($f_\pi$ is fitted to its experimental value)
thus imply $\chi = (178 \, \rm MeV)^4$,
in excellent agreement with the lattice result
$\chi = (175 \pm 5 \, \rm MeV)^4$ of Alles {\it et al.} \cite{Alles:1996nm}.

The mixing angle is then determined to be $\phi = 43.2^\circ$ 
(or equivalently, $\theta = - 11.5^\circ$), for example through the relation
\begin{equation}
\tan2 \phi = \frac{ 2 \sqrt{2} \beta X}{M_{\eta_S}^2-M_{\eta_{NS}}^2} \, ,
\end{equation} 
where 
\begin{equation}
M_{\eta_{NS}}^2 = M_{u\bar u}^2 + 2\beta = M_\pi^2 + 2\beta 
= 0.592 \,\, {\rm GeV}^2 = (769 \,\, {\rm MeV})^2
\end{equation} 
and 
\begin{equation}
M_{\eta_S}^2 = M_{s\bar s}^2 + \beta X^2 = 0.627 \,\, {\rm GeV}^2 
= (792 \,\, {\rm MeV})^2
 \end{equation}
are our {\it calculated} $\eta_{NS}$ and $\eta_S$ masses. They have reasonable
values, in a good agreement with, {\it e.g.}, $\eta_{NS}$ and $\eta_S$ masses 
calculated in the dynamical SU(3) linear $\sigma$ model \cite{Klabucar:2001gr}.

The diagonalization
of the $NS$-$S$ mass matrix gives us the $\eta$ and $\eta'$ masses:
\begin{eqnarray}
        M_{\eta}^2 &=& \cos^2 \phi ~M_{\eta_{NS}}^2    - 
\sqrt{2} \beta X  \sin 2\phi 
       +  \sin^2 \phi ~M_{\eta_{S}}^2 
        \label{eqno99a}
\\
        M_{\eta'}^2 &=& \sin^2 \phi ~M_{\eta_{NS}}^2   +
\sqrt{2} \beta X  \sin 2\phi             
       + \cos^2 \phi ~M_{\eta_{S}}^2~~.
        \label{eqno99b}
\end{eqnarray}
Plugging in the above predictions for $\beta, X, M_{\eta_{NS}}$ and $M_{\eta_S}$,
our model $\eta$ and $\eta'$ masses then turn out to be $ M_{\eta} = 575$ MeV
and $ M_{\eta'} = 943$ MeV. This is in good agreement with the respective
empirical values of 547 MeV and 958 MeV. 

However, the above is not all that can be said about agreement with experiment and other 
approaches. The second thing we may point out is the reasonable agreement
we find if we insert our values of $\beta$, $X$ and $M_{q\bar q'}$'s into our 
model mass matrix and compare it with the $\eta$-$\eta'$ mass matrix obtained 
on lattice by UKQCD collaboration \cite{McNeile:2000hf}.

Third, Ref. \cite{Kekez:2000aw} clearly shows that our approach and results are 
not in conflict, but in fact agree very well with results in the two-mixing-angle scheme 
(reviewed and discussed in, {\it e.g}, Ref. \cite{Feldmann99IJMPA}).
Actually, our results can also be given \cite{Kekez:2000aw,KlKeFB} 
in the two-mixing-angle scheme.

Fourth, what we found from the mass matrix is consistent with what we found 
in the same SD-BS approach through another route, {\it i.e.} from 
$\eta, \eta^\prime \to \gamma\gamma$ processes 
\cite{Klabucar:1997zi,Kekez:2000aw,Klabucar:2000me,Kekez:2001ph}.

The above shows that the consistently coupled SD-BS approach provides a
surprisingly satisfactory description of the $\eta$-$\eta'$ complex, 
especially if one recalls that $\beta$, parameterizing the anomalous 
$\eta_0$ mass shift, was the only new parameter. Namely, all other 
model parameters were fixed already by Ref. \cite{jain93b} providing
the model we used in Refs. \cite{Kekez:1996az,Klabucar:1997zi,Kekez:1998xr,Kekez:1998rw,Kekez:2000aw,Klabucar:2000me,Kekez:2001ph}.
Nevertheless, we would like to point out that even this one parameter, 
$\beta$, can be fixed beforehand. Instead of being a parameter, $\beta$ 
can be obtained through WV formula (\ref{WittenVenez}) from the lattice 
results on the topological susceptibility $\chi$. The central value of 
widely accepted $\chi = (175 \pm 5 \, \rm MeV)^4$ \cite{Alles:1996nm}
would lead to $\beta$ less than 7\% below our model value, which is within upper
error bar anyway. Thus, we can eliminate $\beta$ as a free parameter and
still achieve almost as satisfactory description as the one given above.

\vskip 3mm

\noindent {\bf Acknowledgment:}
D. Klabu\v car thanks the organizers, M. Rosina, B. Golli and S. \v Sirca,
for their hospitality and for the partial support which made possible
his participation at Mini-Workshop Bled 2002.


\end{document}